# Trust and Dependability in Blockchain & AI Based MedIoT Applications: Research Challenges and Future Directions

Ellis Solaiman, Newcastle University, Newcastle upon Tyne, United Kingdom

Christa Awad, Newcastle Hospitals NHS Foundation Trust, United Kingdom

***Abstract—*** *This paper critically reviews the integration of Artificial Intelligence (AI) and blockchain technologies in the context of Medical Internet of Things (MedIoT) applications, where they collectively promise to revolutionize healthcare delivery. By examining current research, we underscore AI's potential in advancing diagnostics and patient care, alongside blockchain's capacity to bolster data security and patient privacy. We focus particularly on the imperative to cultivate trust and ensure reliability within these systems. Our review highlights innovative solutions for managing healthcare data and challenges such as ensuring scalability, maintaining privacy, and promoting ethical practices within the MedIoT domain. We present a vision for integrating AI-driven insights with blockchain security in healthcare, offering a comprehensive review of current research and future directions. We conclude with a set of identified research gaps and propose that addressing these is crucial for achieving the dependable, secure, and patient-centric MedIoT applications of tomorrow.*

**Introduction:** Innovations like the Internet of Things (IoT), Artificial Intelligence (AI), and blockchain are merging with significant potential for modernising digital healthcare applications. The term Medical Internet of Things (MedIoT) has emerged, which aims to boost healthcare efficiency and patient outcomes while reducing costs. But to fully harness these benefits, important challenges must be addressed. Central to these challenges is promoting a culture of reliability and trust. Reliability refers to the consistent performance of technology, delivering expected outcomes without unexpected interruptions or mistakes. Trust, on the other hand, refers to the confidence of users, including healthcare professionals and patients, in the capabilities of the technology to operate correctly, securly, and responsibly when, for example, managing sensitive health data. In this article, we explore how the combination of AI and blockchain can address MedIoT's privacy and trust issues. AI promises to make significant contributions, with algorithms capable of analysing intricate medical data to offer both general and personalised insights into patient health. For example, machine learning models have accurately identified tumours in medical imagery [1]. AI models have also efficiently predicted patient readmissions using electronic health records [2]. These emerging abilities could revolutionise MedIoT applications, leading to tailored medication, enhanced diagnostics, and improved healthcare provision. Similarly, recent blockchain technologies show improved capabilities, offering secure and streamlined health data management. Blockchain's decentralised structure, along with its ability to produce unalterable records, promises superior data security and authenticity. Estonia's e-Health system, which employs blockchain to safeguard patient health data, serves as an example [3]. In MedIoT contexts, blockchain has the potential to ensure a secure setting for data exchange and storage, guaranteeing patient privacy and system reliability. The combination of Blockchain with AI could facilitate real-time health monitoring with enhanced trust and reliability, leading to more effective, patient-focused healthcare. This paper uniquely synthesizes the latest advancements in AI and blockchain in the MedIoT landscape, offering novel insights into their combined potential for enhancing data security, privacy, and patient centered healthcare, and distinctively incorporates both patient and practitioner perspectives to provide a comprehensive, multi-dimentional view of the field. Specifically, this review makes the following contributions: 1) Offers a comprehensive overview of existing research in MedIoT, with a focus on AI and blockchain applications, addressing core issues like data security, privacy, scalability, and ethical concerns, proposing innovative solutions such as energy efficient AI designs and clear regulatory frameworks. 2) Emphasizes the importance of balancing advanced technology with human judgment in healthcare. 3) Identifies ten key challenges and suggests directions for future research to advance the field. 4) Incorporates views from both patients and healthcare practitioners to provide a well rounded understanding of MedIoT's impact and practicality.

## SCENE SETTING

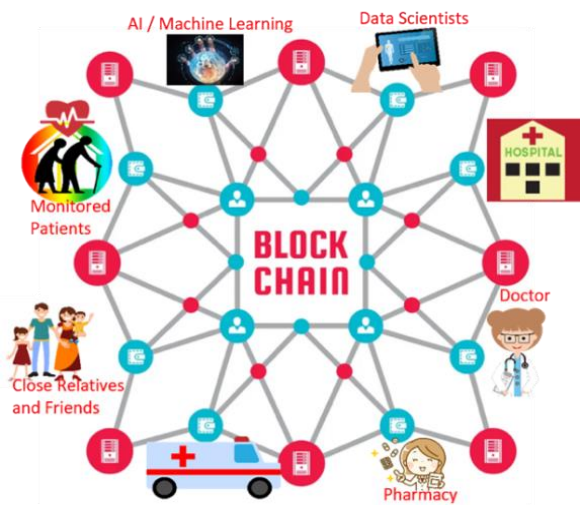

Figure. 1. Combining Blockchain and AI for MedIoT Health Monitoring Applications

Let us envision Mrs. Smith, a 65-year old retiree managing hypertension and Type II diabetes. Traditionally, Mrs. Smith's condition required frequent hospital visits and managing complex records, risking overlooked health fluctuations in between visits. However, in our envisaged digital health future (figure 1), AI and blockchain enhanced MedIoT applications significantly improve her healthcare management. Mrs. Smith uses a smart wristband, acting as her personal health guardian, tracking vital signs 24/7. Blockchain technology secures and renders the wristband's data unalterable. Every heartbeat, and blood sugar level measurement is recorded on a blockchain system, guaranteeing data accuracy, privacy, and easy access. Additionally, smart contracts, a characteristic of blockchain, serve as the custodians of Mrs. Smith's health data. These programmable contracts automatically function when specific pre-set conditions are met, such as data sharing with healthcare providers or alert activation upon anomaly detection. Accompanying this blockchain system is an AI model, a virtual health specialist, designed specifically for Mrs. Smith. This model processes her health data, learns her patterns, and offers personalized health guidance. Upon detecting an unusual surge in her blood sugar levels one day, the system, thanks to the smart contract within the blockchain, immediately notifies Mrs. Smith and her healthcare provider. After thorough data analysis, her provider adjusts her medication, while the AI also suggests dietary and lifestyle modifications based on its comprehensive understanding of Mrs. Smith's health patterns. By enabling proactive health management, AI and blockchain supported MedIoT applications transition patients from being passive beneficiaries to active contributors in their healthcare, thus improving their engagement and experience. Nevertheless, the effectiveness of this system hinges on its reliability and the user's trust. Reliability ensures that the wristband functions correctly, providing precise data, and that he AI accurately provides forecasts and advice. Any malfunction could lead to unnecessary interventions, underlining the requirement for robust MedIoT devices and trustworthy AI models. Trust involves privacy and data security, with potential breaches posing a risk to the system's credibility. The blockchain must guarantee data integrity and requires user consent for data sharing. Finally, for Mrs. Smith to fully trust the system, she needs to understand it, particularly the AI's decision-making process and blockchain's data security protocols. Making these technologies user friendly and comprehensible is essential.

## MED-IOT APPLICATIONS

The combination of AI and blockchain in MedIoT has gained traction due to their collective potential. AI researchers have been working on tailored medicine, real-time health monitoring, and decision support, while blockchain enthusiasts focus on enhancing data security, privacy, and compatibility. In the AI domain, its influence in healthcare is growing. Machine learning and deep learning techniques accurately diagnose diseases from medical imagery. For example, the COVID-CheXNet deep learning algorithm can detect Covid19 from chest X-rays with a 99.99% accuracy rate [4]. Moreover, AI's potential in disease outbreak prediction has been showcased by BlueDot, which predicted the Zika virus spread, later confirmed by the Centres for Disease Control and Prevention [5]. Devices like Fitbit use AI for real-time health insights through heart rate and sleep tracking [6]. In the blockchain space, we see Estonia's e-Health system safeguarding health records [3], MedRec from MIT managing medical data access [7], and the Mediledger project tracking prescription medicines to prevent counterfeits [8]. However, the combination of AI and blockchain in MedIoT offers further avenues for innovation with a growing number of research studies. One study highlights blockchain and AI's role in bolstering healthcare data security and sharing [9]. Another project integrated a Deep Learning model, combining Gated Recurrent Unit (GRU) and Long Short-Term Memory (LSTM) techniques, for distinguishing between Essential Tremor (ET) and Parkinson's Tremor (PST), using blockchain for validation [10]. However, this is a new area of research, and gaps exist. Many AI health models lack clarity, affecting their healthcare uptake [11]. While blockchain promotes data security, it can struggle with healthcare's large data volumes. Most studies on AI and blockchain in MedIoT emphasize data privacy and security, with few exploring other applications like automating processes with AI-driven smart contracts or improving patient interaction using AI on blockchain health records. Implementing AI and blockchain in MedIoT presents many challenges like system and device compatibility, processing vast data in real-time, and ensuring user trust and regulatory adherence. Current research in these sectors is ongoing, suggesting the need for deeper exploration.

# TEN CHALLENGES & FUTURE RESEARCH DIRECTIONS

The following identified challenges are important for enhancing security, ensuring privacy, and improving patient care efficiency. Addressing them will contribute to advancing the technological capabilities of MedIoT, and significantly elevate the quality and reliability of healthcare delivery.

1) **Data Security and Privacy:** MedIoT encompasses a wide network of interconnected devices, raising challenges in data security and privacy. Such devices are vulnerable to malicious attacks, potentially compromising patient data [12]. Blockchain, with its decentralized structure and cryptographic techniques, is seen as a robust solution for data security in this area. Meanwhile, Artificial Intelligence (AI), particularly machine learning, can detect unusual patterns indicating threats. Future research should focus on advanced cybersecurity measures, like Quantum Key Distribution (QKD) which is resistant to quantum computer threats. The exploration of Post-Quantum Cryptography (PQC) is also important, ensuring data protection against quantum machines. Using Homomorphic Encryption in AI, which allows operations on encrypted data without decryption, could be key for data analysis while maintaining privacy. The potential of AI-driven Intrusion Detection Systems (IDS) should be further pursued, helping in pinpointing network anomalies. Additionally, technologies like Zero-knowledge proofs (ZKP), which verify information possession without revealing details, deserve attention, especially when integrated with blockchain, ensuring MedIoT data privacy. Lastly, the potential of blockchain in crafting secure, interoperable Data Marketplaces warrants exploration. Such platforms can promote data sharing for research, enhancing healthcare while ensuring data confidentiality and safety.

2) **Data Integrity and Seamless Interoperability:** Ensuring data trustworthiness and reliability is central to MedIoT applications. The immutable nature of blockchain means that once data is stored, it can't be altered without network consensus, preserving data's authenticity and integrity. For example, patient data from a MedIoT device, once registered on the blockchain, remains consistent and resistant to unauthorized changes. This ensures that such data, crucial for diagnosis or treatment, remains trustworthy. A significant challenge lies in achieving interoperability across various MedIoT systems and devices. Devices, whether they're monitoring heart rates or managing insulin, should communicate and exchange data seamlessly. Future research should focus on devising standardized blockchain protocols and universal AI models to ensure compatibility across multiple systems. This could involve creating protocols allowing different healthcare associated blockchain systems to collaborate easily. AI models should be adaptable to interpret data from any MedIoT device, regardless of its origin or design. By promoting this interoperability, a secure and efficient data exchange can be realized, with blockchain as the backbone and AI for data analysis. Such an approach would boost system trust and dependability, ensuring smooth interactions across various devices and platforms.

3) **Real-time Processing:** In MedIoT, the need for instant data processing in health monitoring applications is challenged by the resource demands of AI, especially deep learning models. For instance, a heart rate monitor must quickly detect anomalies. Due to the vast data and intricate models, real-time processing might be hindered, affecting patient health. A promising solution is edge computing, which processes data near its source, reducing latency and ensuring swift reactions. Imagine a wearable device with an integrated edge-based AI model, allowing immediate on-device data analysis. This would enable instant feedback, crucial during emergencies. Also, developing streamlined energy saving AI models for edge devices could transform real-time data processing. Federated learning, which supports on-device model training with local data, might speed up processing and bolster data privacy [15]. Such techniques could let wearables adapt to individual health patterns directly, yielding rapid, tailored insights while preserving data security. Future research should focus on a hybrid data processing architecture, balancing cloud, edge, and on-device AI models. For example, routine tasks might utilize cloud models, while urgent, critical tasks leverage edge or on-device models. Such designs could offer reliable systems ensuring prompt healthcare actions and efficient resource management.

4) **Scalability:** MedIoT applications incorporating blockchain face scalability challenges due to the resource intensive processes like mining. Imagine a hospital's blockchain-based MedIoT system managing data from hundreds of devices. The sheer volume can impede performance, which can be disastrous when dealing, for example, with heart monitor data. Addressing these issues, research could focus on off-chain and hybrid solutions [14], such as sharding, dividing the blockchain network into smaller segments, each handling its own transactions, thereby improving throughput. Layer 2 solutions, such as Plasma or the Lightning Network, also show promise. These add a layer on top of the main blockchain, facilitating off-chain transaction processing, similar to extra service counters speeding up customer service. Alternative consensus algorithms, like Proof of Stake (PoS) and Delegated Proof of Stake (DPoS), might offer more scalability compared to the common Proof of Work (PoW). PoS can be visualized as an election system prioritizing those with larger stakes to validate transactions, enhancing speed. Similarly, the RAFT (Reliable, Replicated, Redundant, And Fault-Tolerant), and IBFT (Istanbul Byzantine Fault Tolerance) algorithms are significant, with RAFT being akin to a round table where one leader is chosen to make decisions swiftly, while IBFT operates like a diplomatic consensus where multiple trusted parties agree on a transaction to ensure security and speed, even if some participants are unreliable. Technologies tailored for scalability, like Polkadot's multi-chain system, are worth exploration. This system permits simultaneous operation of multiple blockchains, similarly to multi-lane highways boosting traffic flow. Similarly, zk-Rollups, a Layer 2 technique, aggregates several operations into one, improving transaction rates. Future work in these domains could ensure blockchain-based MedIoT systems are both trustworthy and scalable, managing vast data volumes while safeguarding sensitive health data.

5) **User Acceptance and Trust**: For AI and blockchain to thrive in MedIoT applications, user acceptance and trust are crucial. Yet, their inherent complexity and perceived opacity can deter users. A patient might find a blockchain health app daunting and worry about their data's privacy. Addressing these issues requires enhancing system transparency and user friendliness. Future research could focus on creating intuitive interfaces and visualizations to demystify blockchain for users. Similarly, AI models should be crafted to yield clear and relatable results. Systems like Decentralized Identifiers (DIDs) might be explored, acting as tamper-proof digital IDs, improving trust. Moreover, educating users about the technology, its merits, and data security measures can also build trust. Clear explanations, like a doctor clarifying a diagnosis, can enhance understanding and acceptance. Prioritizing such areas can boost the reliability and trustworthiness of MedIoT applications, promoting broader user acceptance.

6) **Explainability in AI:** AI models, especially those using deep learning, often function as "black boxes": they process input data and output results without revealing the details of their operations. This can make it challenging for users to trust such models, especially when clarity is needed, like in health monitoring applications. Explainable AI (XAI) aims to address this by designing AI systems whose decisions are both effective and transparent. Future research could explore methods like Counterfactual Explanations, which clarify AI decisions by showing how different input would lead to varied results. It is like a teacher detailing how each variable influences an answer. Moreover, AI development should consider interdisciplinary collaboration, involving experts from fields like social sciences, law, and ethics. Such an approach can offer users a nuanced, context-specific insight into AI operations, promoting trust and confidence in the technology.

7) **Ethical Considerations:** Protecting the ethical use of sensitive health data is crucial. This involves adhering to data privacy regulations and ensuring patients understand how their information is used, a principle termed; "informed consent". Future research should explore how to integrate these ethical aspects directly into AI and blockchain systems. A key notion is "privacy by design", implying that AI models are constructed with user privacy as a core feature, not an afterthought. Think of it as designing a house with security features from the outset rather than retrofitting them later. Highlighting ethical considerations in MedIoT not only builds trust but also promotes responsible management of health data.

8) **Regulatory Compliance:** A significant challenge, future research should aim to create AI models that inherently align with regulations, thus building user trust. There is also an urgent need for unified international standards to govern AI and blockchain in healthcare, addressing areas like data interoperability, security protocols, and ethical data use. Such guidelines would ensure smooth data exchange between MedIoT systems and mandate strong security measures. Addressing these concerns can help develop trustworthy MedIoT systems that are widely accepted and enhance healthcare results.

9) **Energy Efficiency:** Given that many IoT devices are battery dependent, future research should lean towards devising AI models that are more energy conservative. In terms of deep learning models, techniques such as model pruning, which streamlines neural networks by eliminating superfluous data, and quantization, which refines the numerical precision of parameters, offer ways to diminish computational needs and, consequently, power usage. Additionally, considering alternatives to the traditionally energy-intensive Proof of Work (PoW) mechanism in blockchain, like Proof of Stake (PoS) and Delegated Proof of Stake (DPoS), could further boost energy efficiency [15]. Such energy-efficient strategies can extend the lifespan of IoT device batteries, reducing potential system downtimes, and therefore ensuring more dependable MedIoT applications. This guarantees consistent functionality and promotes user trust by mitigating device failures. In summary, prioritizing energy-efficient AI and blockchain techniques is essential for both environmental benefits and enhancing trust and reliability in MedIoT platforms.

10) **Integration of AI and Blockchain:** The integration of AI and blockchain in MedIoT offers significant promise. For instance, machine learning could be employed to predict and refine blockchain operations, such as optimizing transaction costs by forecasting gas prices on platforms like Ethereum. Another avenue could be enhancing smart contract decision-making with AI, adding an extra layer of automation and insight to blockchain transactions. This would improve the efficiency and reliability of MedIoT applications. Furthermore, blockchain-based systems, when combined with AI, can be tailored for patient engagement and customized healthcare. Here, patient data can be securely held on a blockchain, and AI can then analyze this information to deliver tailored health recommendations, ensuring a bespoke healthcare journey for each user. This dual-tech approach ensures data security while offering custom insights, bolstering patient trust and improving health outcomes.

## IN THE FIELD: PERSPECTIVES FROM PATIENTS AND HEALTH PRACTITIONERS

The integration of wearable technology into patient healthcare management, although not yet a commonplace practice within health services, holds significant potential [16]. Devices like smartwatches and ambulatory blood pressure monitors, despite their current reliance on patient intervention for data transmission, could facilitate a more proactive and personalized approach to health status monitoring. Artificial intelligence (AI), while increasingly prevalent in many sectors, still provokes considerable deliberation within healthcare. Its predictive capabilities, shaped by extensive training on diverse datasets, are undeniably powerful [17]. However, the complexities of medical practice, characterized by a multitude of variables and often subtle nuances, highlight the irreplaceable value of human judgment. Thus, the consensus seems to point towards AI as a supplemental tool rather than an ultimate decision-making authority. Human oversight and clinical intuition remain essential in the diagnostic process and treatment planning, underscoring the necessity for a human-in-the-loop approach when leveraging AI [18]. Emerging technologies like blockchain present opportunities to overhaul the healthcare data ecosystem. This technology could underpin a secure, integrated network for sharing patient data among authorized healthcare providers, although the current data landscape in healthcare systems often falls short of this ideal [19]. Fragmented records and limited accessibility present considerable challenges, underscoring the need for a system where all relevant healthcare professionals can access accurate, comprehensive, and up-to-date patient health records. Security and privacy are cornerstone concerns when considering the integration of such technologies. Rigorous measures are needed to ensure data security and patient privacy, stipulating that only authorized individuals can access sensitive health information and patients maintain control over their health data. The introduction of these technologies also necessitates comprehensive training and awareness among healthcare practitioners. Given the likelihood of resistance, particularly among those less comfortable with technology, implementation strategies should be considered carefully [20]. This might include running parallel systems during transition periods to prevent disruption and foster gradual adaptation. Lastly, it is crucial to resolve health disparities when implementing these technologies. Consideration must be given to factors such as language barriers, cultural differences, and extant disparities in health outcomes between demographic groups. The aspiration should be to leverage these technologies to mitigate health inequalities rather than inadvertently exacerbating them. In conclusion, the integration of technologies such as AI and blockchain into healthcare holds significant potential but also presents substantial challenges. Careful, thoughtful, and inclusive implementation strategies are essential for their successful integration into patient care. Future research should focus on developing protocols for secure and ethical data management, exploring the balance between human judgment and AI in medical decision-making, and assessing the potential of these technologies to address health inequalities. Furthermore, the practicality and acceptability of these technologies from both the practitioner and patient perspectives warrant in depth investigation, with a view towards ensuring a patient centered, equitable healthcare future.

# REFERENCES

1. Shen, L., Wang, Z. J., & Yap, G. A. T. (2019). Artificial Intelligence Versus Clinicians in Disease Diagnosis: Systematic Review. JMIR Medical Informatics, 7(3), e10010.
2. Li, I., Pan, J., Goldwasser, J., Verma, N., Wong, W. P., Nuzumlalı, M. Y., et al. (2022). Neural natural language processing for unstructured data in electronic health records: A review. Comp. Sci. Rev., 46.
3. Semenzin, S., Rozas, D., & Hassan, S. (2022). Blockchain-based application at a governmental level: disruption or illusion? The case of Estonia. Policy and Society, 41(3), 386–401. https://doi.org/10.1093/polsoc/puac014.
4. Al-Waisy, A. S., Al-Fahdawi, S., Mohammed, M. A., et al. (2023). COVID-CheXNet: hybrid deep learning framework for identifying COVID-19 virus in chest X-rays images. Soft Comput, 27, 2657–2672.
5. Rehman, S. U., Shafqat, F., & Niaz, K. (2023). Chapter 16 - Recent artificial intelligence methods and coronaviruses. In Application of Natural Products in SARS-CoV-2. Academic Press, 353-380. ISBN 9780323950473.
6. Mele, C., Spena, T.R., Kaartemo, V., & Marzullo, M.L. (2021). Smart nudging: How cognitive technologies enable choice architectures for value co-creation. J. Busi. Res., 129, 949-960.
7. Azaria, A., Ekblaw, A., Vieira, T., & Lippman, A. (2016). MedRec: Using Blockchain for Medical Data Access and Permission Management. In 2016 2nd International Conference on Open and Big Data (OBD), 25-30.
8. Niu, B., Dong, J., & Liu, Y. (2021). Incentive alignment for blockchain adoption in medicine supply chains. Transportation Research Part E: Logistics and Transportation Review, 152, 102276.
9. Mamoshina, M., Ojomoko, L., Yanovich, Y., Ostrovski, A., Botezatu, A., Prikhodko, et al, (2018). Converging blockchain and next-generation artificial intelligence technologies to decentralize and accelerate biomedical research and healthcare. Oncotarget, 9(5), 5665–5690.
10. Hathaliya, J. J., Modi, H., Gupta, R., & Tanwar, S. (2022). Deep learning and Blockchain-based Essential and Parkinson Tremor Classification Scheme. In IEEE INFOCOM 2022 - IEEE Conference on Computer Communications Workshops (INFOCOM WKSHPS), 1-6.
11. Loh, H.W., Ooi, C.P., Seoni, S., Barua, P.D., Molinari, F., & Acharya, U.R. (2022). Application of explainable artificial intelligence for healthcare: a systematic review of the last decade (2011–2022). Comput. Methods Programs Biomed., 226, 107161. 10.1016/j.cmpb.2022.107161.
12. Rashid, B. (2017). Medical Devices Are the Next Security Nightmare. Wired. https://www.wired.com/2017/03/medical-devices-next-security-nightmare/
13. Short, A. R., Leligou, H. C., & Theocharis, E. (2021). Execution of a Federated Learning process within a smart contract. In 2021 IEEE International Conference on Consumer Electronics (ICCE) (pp. 1-4). IEEE. doi: 10.1109/ICCE50685.2021.9427734.
14. Solaiman, E., Wike, T., & Sfyrakis, I. (2021). Implementation and evaluation of smart contracts using a hybrid on- and off-blockchain architecture. Concurr. Comput. Pract. Exp., 33(1), 1-17.
15. Shi, Z., de Laat, C., Grosso, P., & Zhao, Z. (2023). Integration of Blockchain and Auction Models: A Survey, Some Applications, and Challenges. IEEE Communications Surveys & Tutorials, 25(1), 497-537. doi: 10.1109/COMST.2022.3222403.
16. Sujith, A.V.L.N., Sajja, G.S., Mahalakshmi, V., Nuhmani, S., & Prasanalakshmi, B. (2022). Systematic review of smart health monitoring using deep learning and artificial intelligence. Neurosci. Inform., 2(3), 100028. 10.1016/j.neuri.2021.100028.
17. Alanazi, A. (2022). Using machine learning for healthcare challenges and opportunities. Inform. Med. Unlocked, 30, 100924. 10.1016/j.imu.2022.100924.
18. Topol, E. J. (2019). High-performance medicine: the convergence of human and artificial intelligence. Nature Medicine, 25(1), 44–56.
19. Cerchione, R., Centobelli, P., Riccio, E., Abbate, S., & Oropallo, E. (2022). Blockchain's coming to hospital to digitalize healthcare services: designing a distributed electronic health record ecosystem. Technovation. 10.1016/j.technovation.2022.102480.
20. Greenhalgh, T., Wherton, J., Papoutsi, C., Lynch, J., et al, (2017). Beyond Adoption: A New Framework for Theorizing and Evaluating Nonadoption, Abandonment, and Challenges to the Scale-Up, Spread, and Sustainability of Health and Care Technologies. Journal of Medical Internet Research, 19(11), e367.

**Dr Ellis Solaiman (SFHEA, FBCS)** is a Senior Lecturer / Associate Professor with Newcastle University, UK. His research interests are mainly in the development of trusted distributed smart applications using technologies such as Machine Learning, Blockchain, and Smart Contracts. He is a Fellow of the British Computer Society, and a Senior Fellow of the Higher Education Authority (HEA). Contact him at: ellis.solaiman@ncl.ac.uk.

**Christa Awad, MSc (IP)** is a Lead Integrated Care Clinical Pharmacist and Independent Prescriber currently working at the Integrated Care Pharmacy Hub, Newcastle upon Tyne, UK Hospitals. She has experience working in community pharmacy, primary care and integrated care. Contact her at christa.awad@gmail.com.